
\baselineskip=18pt plus 2pt minus 1pt
\magnification=1200
\overfullrule=0pt
\hsize=5.7truein
\vsize=8.4truein
\voffset=24pt
\hoffset=.1in
\centerline{t-J Model Studied by the New Power-Lanczos Method}
\vskip .3in
\centerline{Y. C. Chen$^1$ and T. K. Lee$^2$ }
\centerline{$^1$Dept. of Physics, National Tsing Hua Univ., Hsinchu, Taiwan}
\vskip .15in
\centerline{$^2$Center for Stochastic Processes in Science and Engineering}
\centerline{and}
\centerline{Department of Physics}
\centerline{Virginia Polytechnic Institute and State University}
\centerline{Blacksburg, VA 24061 USA}
\vskip .15in
\vskip .3in
The initial trial wave function used in a simple
ground-state projection method, the power method, is systematically improved
by using Lanczos algorithm.
Much faster convergence to the ground state
achieved by using these wave functions
significantly reduces the effect of the fermion sign problem. The
energy, spin and charge correlation functions are calculated for the
ground states of the two-dimensional $t-J$ model. Results for
an $8\times8$ cluster with 42 and 26 electrons are presented.
The density correlation function for the $t-J$ model at small J shows
a surprisingly good agreement with that of a system of non-interacting
hard-core bosons.

\vskip .4in

\vfil\eject

Recently we have studied ground-state properties of the $t-J$ model in one$^1$
and two dimensions$^2$ by using a simplified Green function Monte Carlo (GFMC)
method$^3$  - the power method. In this method the ground-state wave
function of a Hamiltonian $H$ is obtained
by applying large powers of the operator $W-H$ to a trial wave function,
where W is a constant. In fermionic systems when the power becomes large
same configurations  with opposite signs will be generated if a
Monte Carlo (MC) algorithm is used. It causes
very large error bars in numerical values. This is
 the famous $\it sign$ problem$^{3,4}$ occurred in MC simulations of
fermionic systems.
In one dimension the phase of the wave function can be fixed to rid of
the sign problem, power method is very successful for all possible
electronic densities$^1$. In two dimensions only at low electronic density
the sign problem is not severe and the converged ground state is obtained$^2$.
At high density
the sign problem makes the power method ineffective to study this
interesting region for high-temperature superconductors.

The freedom to choose the trial wave function is
one of the special properties of the ground-state projection method.
A trial function chosen inappropriately would require
a lot of computer time to converge to ground state. Sometimes the sign problem
 makes the convergence impossible.
It is imperative
to have a good trial function to reduce the number
of negative terms which increases with the power.

In the last several years variational MC method has been widely
used to study the $t-J$ model$^{5,6.7}$. Several innovative wave
functions have been proposed for the ground state.
Some of them tested by the power method are not as close to the ground state
as one would have anticipated.
There were few methods that we can use
to systematically improve the trial wave function. Recently Heeb and Rice$^8$
proposed to use Lanczos$^9$ iteration to obtain better wave functions.
The effectiveness of the method is demonstrated
by studying the two-dimensional antiferromagnetic Heisenberg model.
A few years earlier, Caffarel et al$^{10}$ have also used a variation
of the Lanczos algorithm to study $L_iH$ molecule.


Although Lanczos method$^9$ is best known in searching for wave functions of
small clusters,
the method itself is quite general. Starting with a wave function
$\left| \phi_0 \right>$, we can generate a tri-diagonal matrix by using
the recurrence relation$^9$
$$   H \left| \phi_0 \right> = a_0\left| \phi_0 \right> + b_1\left| \phi_1
\right>
$$
$$
H \left| \phi_n \right> = a_n\left| \phi_n \right> + b_n\left| \phi_{n-1}
\right>
+b_{n+1} \left| \phi_{n+1} \right> \eqno(1)
$$
where $n=$1,2,..etc.
The matrix elements, $a_n$ and $b_n$, are related to
the moments of the Hamiltonian. For example,
$a_0=\left< \phi_0 \left| H \right| \phi_0 \right>$ and
$b_1=\sqrt{ \left< \phi_0 \left| (H-a_0)^2 \right| \phi_0 \right>  }$.
When $n$ increases, the lowest
eigenvalue of the tri-diagonal matrix approaches the ground-state energy.
And the eigenfunction of this lowest eigenvalue gets closer to the
ground-state wave function.
It is straight forward to show that in succesive iteration
 the eigenstates have the form
$$\left| \Psi_1 \right>= \left| \phi_0 \right> + C_1 {1\over N}H \left|
\phi_0 \right>, \eqno(2)
$$ and
$$\left| \Psi_2 \right>= \left| \phi_0 \right> + C'_1 {1\over N}H \left|
\phi_0 \right> +C'_2 {1\over N^2}H^2 \left|\phi_0 \right>, \eqno(3)
$$
... etc. These functions form the
basis in Krylov subspace$^{11}$. The C's are calculated from the
the matrix elements, $a_n$ and $b_n$, by diagonalizing the matrix.

Heeb and Rice$^8$ propose to calculate the matrix elements, $a_n$ and
$b_n$, by using the Monte Carlo technique. The C's are then determined.
However, in this method the values of
the matrix elements must be calculated very accurately. A small
error will produce large uncertainty in the eigenvalues and in C's.
Here we choose an alternative.
We treat C's as the variational parameters.
The wave function with the optimal energy is the
eigenfunction with the lowest eigenvalue. This is more efficient and sometimes
 more accurate than diagonalizing the matrix$^{8,10}$.

The result of this variational Lanczos algorithm is that
we have a sequence of wave functions,
$\left| \phi_0 \right>,\left| \Psi_1 \right>,\left| \Psi_2 \right>$,.. etc.,
with lower and lower energy. Besides the statistical fluctuation
associated with the MC technique, the same result
as the Lanczos method will be obtained.
The fact that this method does not need very large
memory space to store all the configurations as in the usual Lanczos method
is one of its biggest advantages.
But there is a practical difficulty with this approach of getting the
ground state.
Each time the Hamiltonian $H$ is applied
to a particular configuration the number of new configurations generated is
of order of, N, the size of the cluster. It is impractical to
do any calculation with $\left| \Psi_n \right>$ for $n\geq3$
for a cluster of 64 sites or greater.
A more efficient way to obtain the ground state is to use
$\left| \Psi_1 \right>$ or $\left| \Psi_2 \right>$ as
the trial wave functions in the power method.
 We shall refer to this as the power-Lanczos (PL)
method. If the starting trial function before the power method is applied is
$\left|\Psi_n\right>$ we shall call it PL$n$. PL0 is the same as the usual
power method. For the reason discussed above  we shall only consider
PL1 and PL2 in this paper.


Once the optimal wave functions  $\left| \Psi_1 \right>$ and
$\left| \Psi_2 \right>$
are determined, we can proceed to calculate quantities such as
${\left< \Psi_1 \left| (W-H)^p \right| \Psi_1 \right> \over
\left< \Psi_1 | \Psi_1 \right>} $, where $p$ is the power.
It is sufficient to choose the constant W to be zero in the $t-J$ model.
The procedure to carry out this
part is the same as the power method$^1$.

We use several different forms of $ \left| \phi_0 \right>$ to study
the $t-J$ Hamiltonian. The familiar
Gutzwiller wave function (GWF)$^{12}$ is just the wave function for an ideal
Fermi
gas excluding configurations with doubly occupied sites.
Another function proposed by Hellberg and Mele$^6$ and used
by Valenti and Gros$^7$ in 2D was shown to be close to the ground state
at low density$^2$. This function, which we shall call HMVG, is basically
of the same form as GWF, i.e. a Slater determinant for up-spin electrons and
one for down-spin electrons. In addition to these two determinants, it has a
long range correlation part between all the particles,
$\Pi_{i<j} \left| {\bf r}_i-{\bf r}_j \right |^{\nu}$ ( while for
nearest-neighbor particles we choose $\nu=0$). Besides these two functions
we also use the projected BCS state or the resonating-valence-bond
state$^{5,13}$
with either s-wave or d-wave symmetry for the gap order parameter.

 The energy,
$E=\left<H^{2p+1}\right>/\left<H^{2p}\right>$, as a function of power p
is plotted in Fig.1 for 10 particles in
a 4$\times$4 lattice.
Here we consider $J=2t$ and GWF is chosen to be the initial trial
function $ \left| \phi_0 \right>$. The open triangles
are the result of Lanczos algorithm for different orders of
iteration. These results are obtained exactly using the usual Lanczos method
described briefly in equation 1. The variational energy of GWF is about
$5\%$ above the ground state energy.
This difference is reduced to about $0.3\%$ by using the
second order wave function. The solid circles, squares and triangles are the
results of PL0, PL1 and PL2 by using
$\left| \phi_0 \right>,\left| \Psi_1 \right>$, and $\left| \Psi_2 \right>$
respectively. For $\left| \Psi_1 \right>$ of equation (2), we choose
$C_1$ to be $0.8$. We have $C'_1=1.72$ and $C'_2=0.72$ in
$\left| \Psi_2 \right>$ of equation (3). Clearly, when the power becomes
large enough, all these three algorithms would produce ground-state energy.
For comparison, we also calculated the energy exactly without using the
MC technique in PL0, PL1 and PL2.
They are shown by the dashed lines. The excellent agreement between exact and
MC calculations reaffirms  the stability of MC technique.

In Fig. 1 we note the relatively large error bars at powers greater than 4.
This is mainly due to the fermion sign problem. The effect of this sign is
studied by calculating the ratio of contributions from the negative terms and
contributions from the positive terms in the quantities $\left<(-H)^m\right>$.
In the inset of figure 1 this ratio is plotted as a function of m. This
ratio is about the same for different PL's.
At power equal
to 6 or $m=13$, the negative terms are as large as seventy
percent of the positive terms.
It is very time consuming to get good statistics.

 The data for PL0 and PL1 are
obtained by averaging ten to twenty independent groups. Each group usually
consists of one to two thousand starting configurations.
Each starting configuration would produce several hundred
terms in the evaluation of powers of $H$. At large power we usually need to
have more runs to reduce the fluctuations caused by the sign problem.
The calculations are all done in workstations like HP720.
For PL0, the longest calculation is about six hours for this 4$\times$4
cluster. It takes about four to six times longer for PL1. If we had used the
same
number of runs for PL2 we would probably increase the CPU time  by another
factor of four to six. Instead, we reduced the number of runs in PL2,
therefore we have a somewhat larger error at power equal to four.
The amount of computer time quoted above is only for a small cluster of
size 4$\times$4. We estimate the time needed for an 8$\times$8 cluster
with same electronic density is about ten times longer. Clearly, it is
quite impractical to carry out a calculation for large powers in PL2.
Hence, below we shall present mostly results of PL0 and PL1 for 8$\times$8
clusters and may be a few results of PL2 without powers.

 In figure 2(a) and 3(a) energy as a function of
power is plotted for 26 and 42 electrons repectively,
in an $8\times8$ lattice for $J=0.1t$.
VGHM function with $\nu=0.04$ is used as  $\left| \phi_0 \right>$ and its
variational energy is about $3\%$ above the ground state energy for
$\left<n\right>={26 \over 64}$, but more than $5\%$ for
$\left<n\right>={42 \over 64}$. The situation is improved substantially
in PL1 when  $\left| \Psi_1 \right>$ of equation (2) is used. $C_1=1.33$
in 2(a) and 1.49 in 3(a).
In figure 2(a) the large error bars at
power equal to 6 and 8 makes it difficult to determine
the exact ground state energy. The effect of the negative sign
is larger when the density is increased from $\left<n\right>={26 \over 64}$
to ${42 \over 64}$. Neverthless,
it seems quite reasonable to conclude that the ground-state
energy should be within half a percent of the variational energy of
$\left| \Psi_2 \right>$  which are represented by the solid
triangles. For PL2 in Fig.2, $C'_1=2.66$ and $C'_2=1.77$, and $C'_1=2.95$
and $C'_2=2.18$ in Fig.3.

Besides the energy we also calculate the equal time correlation functions, in
particular, the spin and density structure factors,
$S({\bf k})$ and $N({\bf k})$ respectively.
These structure factors are plotted along
$\Gamma$-X-M--$\Gamma$
direction in the Brillouin zone in figures 2(b) and 2(c) for
$\left<n\right>={26 \over 64}$ and in figures 3(b) and 3(c) for
$\left<n\right>={42 \over 64}$.
 Open circles
represent the variational result of VGHM$_{\nu=0.04}$,
and open squares are for PL1 without power.
Open triangles are results of
PL1 with power equal to 6 for $\left<n\right>={26 \over 64}$ and power
equal to 5 for $\left<n\right>={42 \over 64}$. The solid lines connecting
triangles are guides for the eyes.
We note that
the results changed markedly between the initial variational
wave function and the first order Lanczos wave function.
The situation seems to get worse when the density $\left<n\right>$ increases,
even though VGHM wave function still has the best variational energy
at $J=0.1t$.
This points out a possible deficiency in using the trial
wave function VGHM to understand ground states of the $t-J$ model at high
electronic density. For comparison, we also show the results of
GWF  as dotted lines. GWF clearly does not
reflect the correlation of the ground state.
So far we have not yet found a
wave function that would have energy within $5\%$ of the ground state.

Recently we have shown that many results of the $t-J$ model at
low electronic density are qualitatively consistent with the prediction of
the Tomonaga-Luttinger liquid$^{14}$ in one dimension. The cusps or peaks
at ${\bf k}=2{\bf k}_F$ in $S({\bf k})$ are enhanced over the variational
results of VGHM. $N({\bf k})$ has a maximum at ${\bf k}=(\pi,\pi)$.
But we cannot$^{15}$
identify in $N({\bf k})$ the
characteristic wave vector $2{\bf k}_F^{SF}$ associated with
spinless fermions (SF) as claimed by Putikka et al.$^{16}$ using the
high temperature expansion. Here we try to understand $N({\bf k})$
from a different point of view.

One way to treat the constraint of no double occupancy in the $t-J$ model
is to write the fermion operator as a product of a hard-core boson and a
fermion
operator. While the fermion operator represents spin degree of freedom,
the boson is for charge degree of freedom. This is the so called
slave-boson approach$^{17}$.
If the separation of charge and spin indeed occurs it will be
most apparent in the limit of vanishing $J$ where
the dynamics is controlled by the charge hopping or the motion of the
hard-core bosons.  We may expect the charge correlation to
be simialr to that of a system of non-interacting hard-core bosons.
In one dimension hard-core bosons and spinless fermions are
equivalent, but they are not in two dimensions. A careful examination of
the correlation function of the hard-core bosons in a 2D lattice is needed.

We have
calculated the ground-state correlation function of a system
of non-interacting hard-core bosons by using the power method. The trial
wave function is of the form of Jastrow type described in Ref. 18. Details
of this calculation will be presented elsewhere.
Results of density correlation are represented by the
solid circles in figure 2(c) and 3(c).
They almost lie exactly on top of the
triangles representing the result of PL1, except$^{19}$ at very small ${\bf
k}$.
A similar result$^{20}$ has been found for the infinite-U Hubbard model for
small clusters.
 On the other hand, $N({\bf k})$
of SF as shown by the dashed lines in figure 2(c) and 3(c)
is not as close to the result of $t-J$ model.
The fact that hard-core bosons have almost the same density-density correlation
as the charges in the $t-J$ model does not by itself prove the separation
of spin and charge. But this and other evidences$^{2,15,16}$
make the idea$^{21}$ of separation of charge and spin in the $t-J$ model much
more
plausible.

In summary, we have presented a new modified power method using
a systematically improved trial wave function obtained by the Lanczos method.
Even the wave function obtained by a first order iteration
greatly improves the rate of convergence to the ground state.
This faster convergence
significantly reduces the effect of the fermion sign problem that has
plagued fermion MC calculations so far.
For the first time very accurate ground-state results are obtained
for electronic density as high as $65\%$ in an $8\times8$ cluster.
A surprising result has been found.
The density-density correlation obtained at small J
is very close to that of a system of
non-interacting hard-core bosons.

The power-Lanczos method presented above is in principle an exact approach to
obtain the ground state.
Unlike the fixed-node method$^{22}$, the results are not overwhelmingly
influenced by the initial choice of trial function. The numerical
algorithm we have used is the simplest among many complicated GFMC methods.
It is already sufficient to get very accurate results. More sophisticated
approaches, such as using guiding function$^{3,22}$, would be explored
in the future. Calculations for much larger clusters
are feasible now.

\medskip
TKL would like to thank Materials Science Center and Department of
Physics of National Tsing Hua University for their hospitality
during his visit where part of this work is carried out.
This work was
partially supported  by
the National Science Council of Republic of China, Grant Nos.
NSC83-0511-M007-004.
\medskip

\vfil\eject

\centerline{REFERENCES}
\medskip
\item{1}Y.C. Chen and T.K. Lee, Phys. Rev. B\underbar{47}, 11548
(1993); {\it Proceedings of the Beijing International
Conference on High-Temperature Superconductors}, edited by Z.Z. Gan, S.S. Xie,
and Z.X. Zhao, 829 (1993), World Scientific, Singapore.

\item{2}Y.C. Chen and T.K. Lee, to appear in Z. Phys. B.

\item{3}{\it Monte Carlo Methods in
Statistical Physics}, edited by K. Binder (Springer-Verlag, Berlin),
1977 and {\it ibid.} Vol. II (1984).

\item{4}Shiwei Zhang and M.H. Kalos, Phys. Rev. Lett. \underbar{67},
3074 (1991); S. B. Fahy and D. R. Hamann, Phys. Rev. Lett.
\underbar{65}, 3437 (1990).

\item{5}T.K. Lee and Shiping Feng, Phys. Rev. B\underbar{38}, 11809 (1988);
T.K. Lee and L.N. Chang, Phys. Rev. B\underbar{42}, 8720 (1990).

\item{6}C. Stephen Hellberg and E.J. Mele, Phys. Rev. Lett. \underbar{67},
2080 (1991).

\item{7}R. Valenti and C. Gros, Phys. Rev. Lett. \underbar{68}, 2402
(1992).

\item{8}E.S. Heeb and T.M. Rice, Z. Phys. B\underbar{90}, 73 (1993).

\item{9}R. Haydock, The Recursive Solution in The Schr$\ddot{o}$dinger
Equation, in {\it Solid State Physics}, vol. 35, 215 (Academic Press, 1980).

\item{10}M. Caffarel, F.X. Gadea and D.M. Ceperley, Europhys. Lett.
\underbar{16}, 249 (1991).

\item{11}B.N. Parlett, {\it The Symmetric Eigenvalue Problem},
(Prentice-Hall), Seires in Computational Mathematics, 1980.

\item{12}C. Gros, R. Joynt, and T.M. Rice, Phys. Rev. B\underbar{36}, 381
(1987).

\item{13}C. Gros, Phys. Rev. B\underbar{38}, 931 (1988).

\item{14}S. Tomonaga, Prog. Theo. Phys., \underbar{5}, 544 (1950);
J.M. Luttinger, J. Math. Phys., \underbar{4}, 1154, (1963).

\item{15}Y.C. Chen, A. Moreo, F. Ortolani, E. Dagotto, and T.K. Lee,
unpublished.

\item{16}W.O. Putikka, R.L. Glenister, R.R.P. Singh and H. Tsunetsugu,
unpublished.

\item{17}P.A. Lee and N. Nagaosa, Phys. Rev. B\underbar{46}, 5621 (1992).

\item{18}L. Reatto and G.V. Chester, Phys. Rev. \underbar{155}, 88 (1967);
A.A. Ovchinnikov, unpublished.

\item{19}Larger power is needed to get
convergence of the correlation function at small ${\bf k}$
or at distance of order of the size of the cluster.

\item{20}M.W. Long and X. Zotos, Phys. Rev. B\underbar{48}, 317 (1993-I).

\item{21}P.W. Anderson, Phys. Rev. Lett. \underbar{64}, 1839 (1990).

\item{22}H.J.M. van Bemmel, D.F.B. ten Haaf, W. van Saarloos, J.M.J. van
Leeuwen and G. An, Phys. Rev. Lett. \underbar{72}, 2442 (1994).

\vfil\eject

\centerline{Figure Captions:}
\medskip
\item{Fig. 1} Energy as a function of power for 10 electrons in a
$4\times4$ cluster. GWF is the trial function used for $J=2t$.
The solid circles, squares and triangles represent results for
PLO, PL1 and PL2, respectively. The dashed lines represent exact
results without using Monte Carlo technique. Open triangles are the
exact results obtained from each order of Lanczos iteration.
In the inset ratio of contributions from negative terms and
contributions from positive terms
as a  function of the power m in $\left<(-H)^m\right>$.

\item{Fig. 2} (a)Energy as a function of power for
$\left<n\right>={26 \over 64}$ and $J=0.1t$  calculated using
 PL0, PL1 and PL2 algorithms. $C_1$=1.33 in $\left| \Psi_1 \right>$,
$C'_1=2.66$ and $C'_2=1.77$ in $\left| \Psi_2 \right>$.
(b) spin structure
factor S($\bf k$) and (c) density structure factor N($\bf k$)
in the $k$ space along $\Gamma$-X-M-$\Gamma$ directions.
Empty circles represent variational results using VGHM function with
$\nu=0.04$. Open squares are results of PL1 without power. Open
triangles are PL1 results at power equal to 6. Solid circles
represent results of non-interacting hard-core bosons. Dotted line is the
variational results of GWF. Results of SF are represented by
the dashed line.

\item{Fig. 3} (a)Energy as a function of power for
$\left<n\right>={42 \over 64}$ and $J=0.1t$  calculated using
 PL0, PL1 and PL2 algorithms. $C_1$=1.49 in $\left| \Psi_1 \right>$,
$C'_1=2.95$ and $C'_2=2.18$ in $\left| \Psi_2 \right>$.
(b) spin structure
factor S($\bf k$) and (c) density structure factor N($\bf k$)
in the $k$ space along $\Gamma$-X-M-$\Gamma$ directions.
Empty circles represent variational results using VGHM function with
$\nu=0.04$. Open squares are results of PL1 without power. Open
triangles are PL1 results at power equal to 5. Solid circles
represent results of non-interacting hard-core bosons. Dotted line is the
variational results of GWF. Results of SF are represented by
the dashed line.

\end